\documentclass{pramana}

\usepackage{graphicx,amsmath,bm}

\usepackage{caption}
\usepackage{subcaption}
\begin{document}

\title{Nonlinearity in Data with Gaps: Application to Ecological and Meteorological Datasets}


\author{Sandip V. George\textsuperscript{1}, G. Ambika\textsuperscript{1,*}}
\affilOne{\textsuperscript{1} Indian Institute of Science Education and Research, Pune-411008, India\\}


\twocolumn[{

\maketitle

\corres{g.ambika@iiserpune.ac.in}


\begin{abstract}
Datagaps are ubiquitous in real world observational data. Quantifying nonlinearity in data having gaps can be challenging. Reported research points out that interpolation can affect nonlinear quantifiers adversely, artificially introducing signatures of nonlinearity where none exist. In this paper we attempt to quantify the effect that datagaps have on the multifractal spectrum ($f(\alpha)$), in the absence of interpolation. We identify tolerable gap ranges, where the measures defining the $f(\alpha)$ curve do not show considerable deviation from the evenly sampled case. We apply this to the multifractal spectra of multiple data-sets with missing data from the SMEAR database. The datasets we consider include ecological datasets from SMEAR I, namely  CO$_2$ exchange variation, photosynthetically active radiation levels and soil moisture variation time series, and meteorological datasets from SMEAR II, namely dew point variation and air temperature variation. We could establish multifractality due to deterministic nonlinearity in two of these datasets, even in the presence of gaps.
\end{abstract}



}]


\year{2017}
\setcounter{page}{1}
\lp{8}

\section{Introduction}
The presence of gaps in observational data is one of the major problems that affect real world datasets. Such gaps or the consequent uneven sampling in time series can be a major impediment in detecting nonlinearity in them. The most common way to deal with datagaps has been to interpolate through them to make an evenly sampled dataset. Past studies have shown that interpolation can have adverse effects on nonlinear time series quantifiers\cite{gra86}. Some recent studies have been conducted on the effect datagaps have on nonlinear quantifiers when the analyzed time series is not interpolated through, which showed that there are certain regions of gap size and frequency where correlation dimension values are affected more\cite{geo15}. We propose to extend these studies to the multifractal spectrum. Such an analysis helps us identify how susceptible the multifractal spectrum is to gaps in data. 

Datagaps are common in real world datasets. Astrophysical data, ecological data, meteorological data etc are often plagued by gaps\cite{buc95, zliobaite2014regression}. They may occur due to multiple reasons like instrumental failure, cloud cover, adverse weather conditions etc. Most nonlinear quantifiers require large evenly sampled datasets in order to make predictions\cite{eckmann1992LimitationOndDimensions,jeffries1998bicoherenceElectrical}. One can exploit the presence of multiple smaller evenly sampled segments in order to calculate quantifiers like bicoherence\cite{geo17}. However no analysis has been done so far on the effect of datagaps on the multifractal spectra. Such an analysis is extremely important in the context of quantifying the multifractal nature of the large number of datasets that come with missing data. 

The multifractal spectrum, $f(\alpha)$, is a detailed characterization of the fractal structure of the system in a reconstructed phase space. Unlike an average measure like the correlation dimension, $D_2$, the $f(\alpha)$ spectrum takes into consideration the local contributions of different regions of the attractor. The idea of $f(\alpha)$ spectrum can be applied beyond attractors in phase space, to describe many natural objects\cite{hil00}. A complete characterization of the multifractal curve was shown to be achieved with a set of four parameters\cite{har09mul}. In this study we aim to check how much variation these parameters show when the time series used for embedding contains missing data.

To determine the susceptibility or resilience of the $f(\alpha)$ spectrum to datagaps, we first start with an evenly sampled time series of a standard system. We subject the evenly sampled time series to multifractal analysis by using the algorithm described in \cite{har09mul}. We then remove data from the evenly sampled time series, progressively. The size of the data removed and frequency of the data removal are drawn from two Gaussian distributions. The time series with datagaps are also subject to multifractal analysis. The deviation of the multifractal quantifiers like the width of the spectrum, from the evenly sampled value is treated as an index of susceptibility of that quantifier to datagaps. We can hence identify the regions of gap size and gap frequency where the quantifiers can give reliable conclusions from the nature of the $f(\alpha)$ curve.

As an illustration, we proceed to analyze real world datasets with missing data. The datasets we consider are obtained from the Station for Measuring Forest \\Ecosystem-Atmosphere Relation (SMEAR) in Finland \cite{hari2013station}. We consider ecological and meteorological datasets obtained from SMEAR I and SMEAR II stations respectively. The ecological datasets we consider are related to photosynthesis levels in pine trees in the V{\"a}rri{\"o} forest. We look into the multifractal spectrum of CO$_2$ exchange in pine shoots, which acts as a proxy for photosynthesis. We also obtain the $f(\alpha)$ curves for the time series of levels of photosynthetically active radiation and soil moisture, both of which are known to influence photosynthesis rates\cite{Farquhar1980BiochemicalModelPhotosynthesis, shimshi1963soilmoisture}. For the meteorological datasets we consider the time series of air temperature and dew point, obtained from the SMEAR II station in the Hyyti{\"a}l{\"a} forest. Both the time series give a continuous $f(\alpha)$ curve, indicative of multifractality in its phase space structure. 
\section{Analysis of synthetic datasets}
In this section we consider the effect of datagaps on the structure of the $f(\alpha)$ curves for standard nonlinear systems like R{\"o}ssler and Lorenz systems. We start from large evenly sampled datasets of the $x$ variable of these systems and introduce gaps into the data. Since the gaps arise from multiple independent sources, we take that both the position and size of the gaps follow Gaussian distributions,  
\begin{equation}
G_S(s;m_s,\omega_s) = \frac{1}{{\omega_s \sqrt {2\pi } }}e^{{{ - \left( {s - m_s } \right)^2 } \mathord{\left/ {\vphantom {{ - \left( {x - m_s } \right)^2 } {2\omega_s ^2 }}} \right. \kern-\nulldelimiterspace} {2\omega_s ^2 }}}
\end{equation}
\begin{equation}
G_P(p;m_p,\omega_p) = \frac{1}{{\omega_p \sqrt {2\pi } }}e^{{{ - \left( {p - m_p } \right)^2 } \mathord{\left/ {\vphantom {{ - \left( {x - m_p } \right)^2 } {2\omega_p ^2 }}} \right. \kern-\nulldelimiterspace} {2\omega_p ^2 }}}
\end{equation}
$G_S$ with mean, $m_s$ and standard deviation, $\omega_s$  determines the size of the gap. Gaussian, $G_P$ with mean $m_p$ and standard deviation $\omega_p$ determines the position of the gap.

Starting from any point in the time series, $G_P$ determines the position of the next gap, i.e. it determines the extend of gapless data. Hence it is an inverse measure of frequency of gaps. $G_S$ determines how big the gap is at a particular position.

\subsection*{Effect of datagaps on $f(\alpha)$ spectrum}
A detailed characterization of the non uniformities of the attractor is captured by the generalized dimensions, $D_q$ \cite{hil00,har09mul}. 
In an analogous formulation, if one covers the attractor with boxes of size R, the probability $p_i$ for points to fall inside the $i^{th}$ box, is found to scale as
\begin{equation}
    p_i(R)=R^{\alpha_i(R)}
\end{equation}
Then the number of such boxes that have $\alpha$ between $\alpha$ and $\alpha+\Delta\alpha$ is\cite{Mogens1985GlobalUniversalityFalp}
\begin{equation}
    n(\alpha,R) \propto R^{-f(\alpha)}
\end{equation}
$D_q$ and $f(\alpha$) are related through Legendre transformations. This provides a method to compute the $f(\alpha)$ spectrum numerically from the $D_q$ values. We can characterize the $f(\alpha)$ curve using the following function fit
\begin{equation}
f(\alpha)=A(\alpha-\alpha_{1})^{\gamma_1} (\alpha_{2}-\alpha)^{\gamma_2}
\label{eqn:fal}
\end{equation}
and $\alpha_1$, $\alpha_2$, $\gamma_1$ and $\gamma_2$ serve as unique parameters that can characterize the $f(\alpha)$ completely\cite{har09mul}. The $f(\alpha)$ spectrum finds widespread use across a variety of different fields\cite{de1999multifractal,ivanov1999multifractality, kavasseri2005multifractal, harikrishnan2011nonlinear}. 

\begin{figure}[ht]
\centering
\includegraphics[width=.45\textwidth]{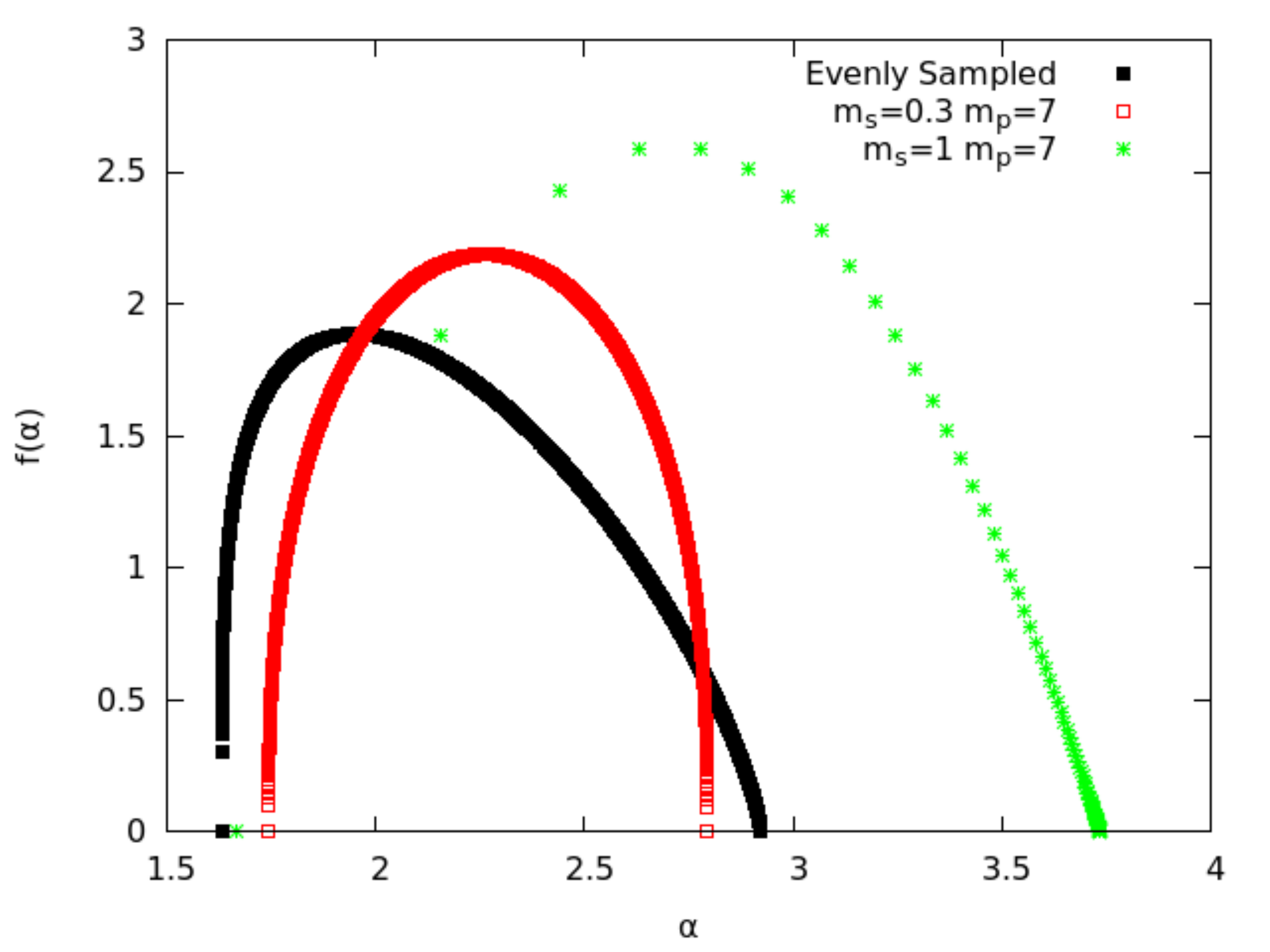}
\caption{\label{fig:falvarross} Variation of $f(\alpha)$ curves for R{\"o}ssler for different values of $m_s$ and $m_p$. We see that while the width remains broadly same as evenly sampled (black) case and for $m_s$ = .3$\tau$, $m_p$ = 7$\tau$, the overall shape changes. The width  and overall shape change drastically for the case of $m_s$ = 1$\tau$, $m_p$ = 7$\tau$. }
\end{figure}

\begin{figure}[hp]

\begin{subfigure}{\textwidth}
\includegraphics[width=.45\textwidth]{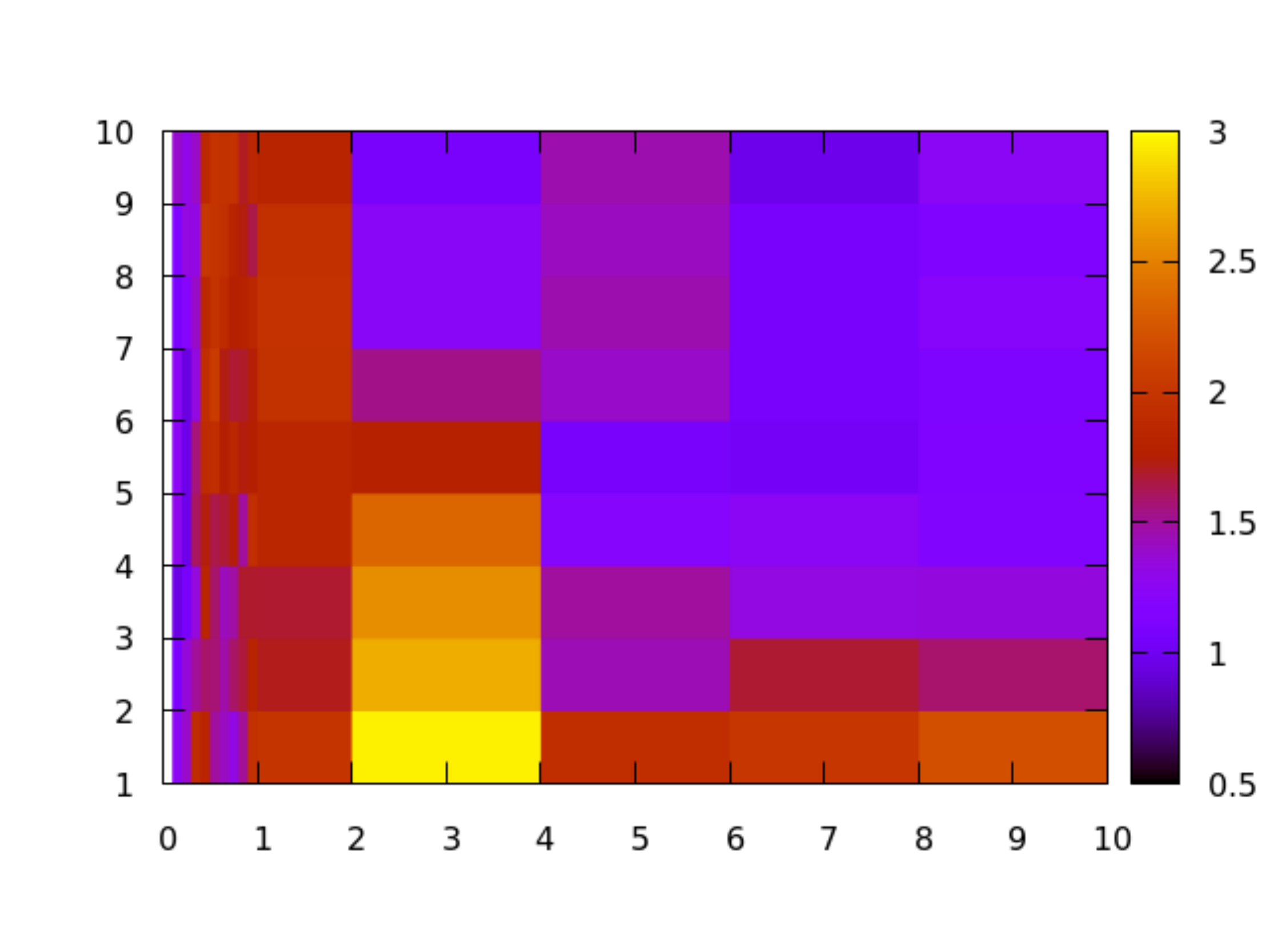}
\subcaption{\label{fig:falross}}
\end{subfigure}
\begin{subfigure}{\textwidth}

\includegraphics[width=.45\textwidth]{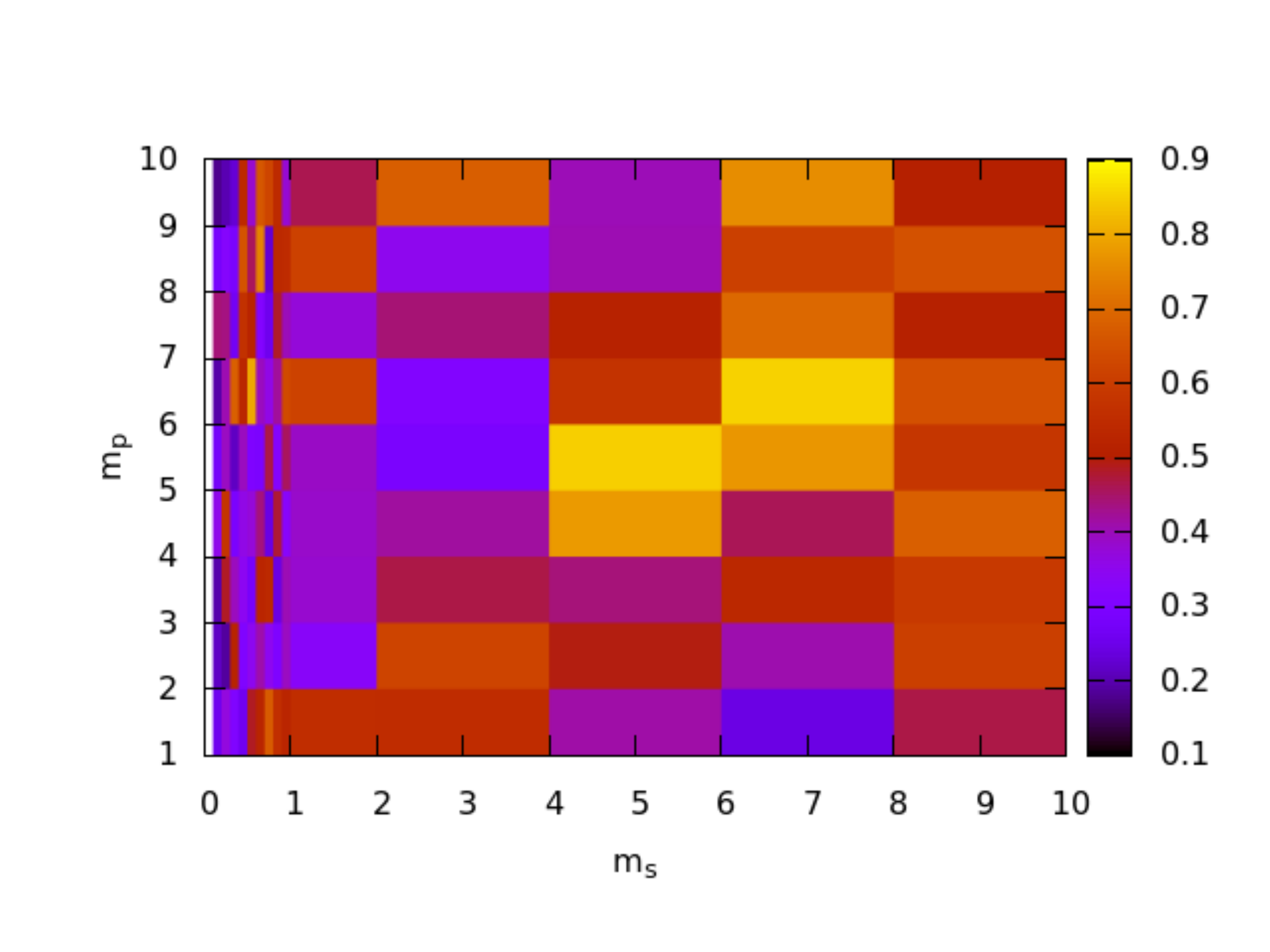}

\subcaption{\label{fig:gam1ross}}

\end{subfigure}
\begin{subfigure}{\textwidth}
\includegraphics[width=.45\textwidth]{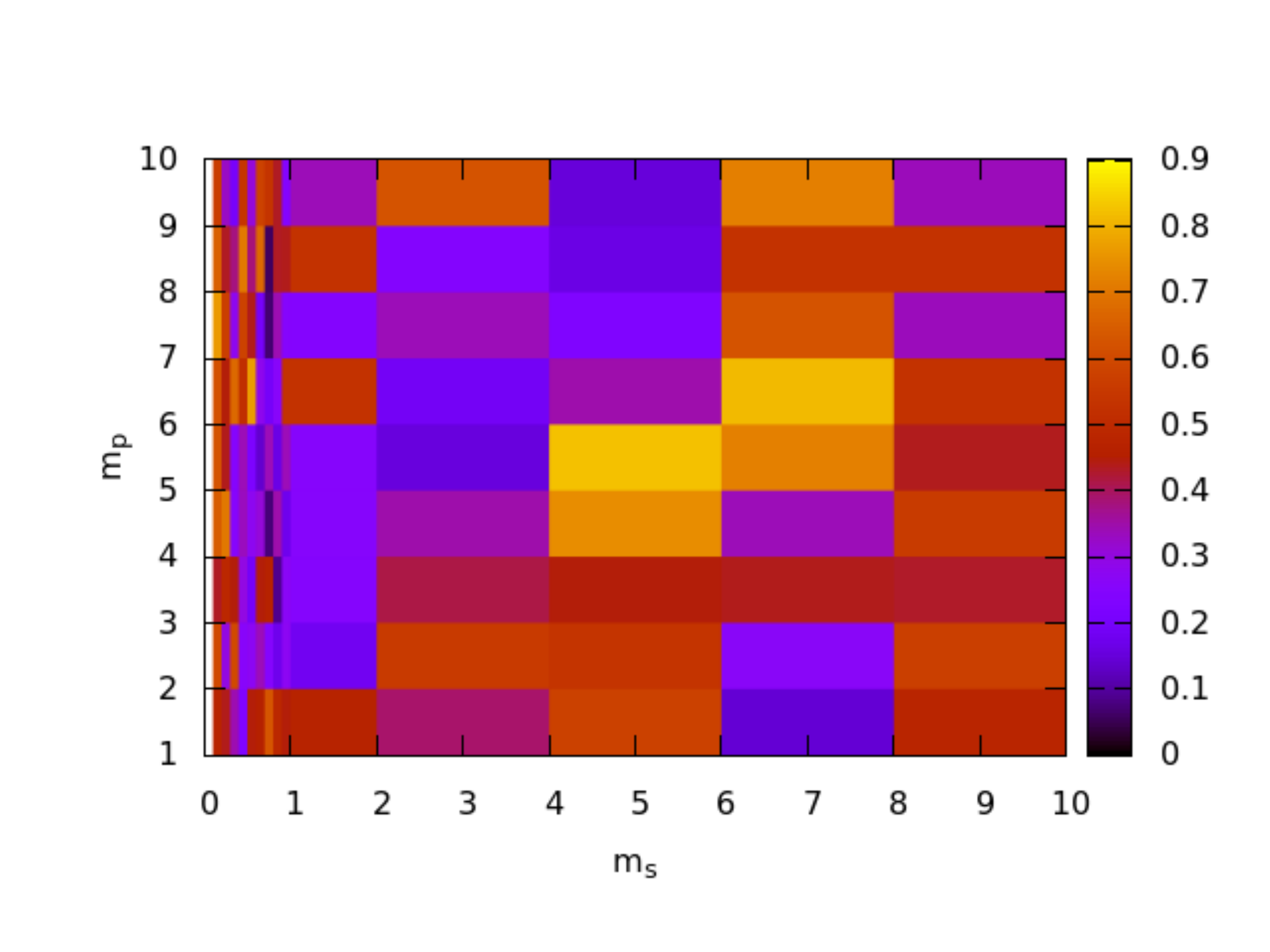}
\subcaption{\label{fig:gam2ross}}
\end{subfigure}
\caption{\label{fig:alpvar} Variation of (a) $\delta \Delta\alpha^{rel}$ (b)$\delta\gamma^{rel}_1$ and (c) $\delta\gamma^{rel}_2$ for varying gap size and frequency for the R{\"o}ssler. In (a) we see a region around 1$\tau$ where the  values deviates greatly from the evenly sampled value for R{\"o}ssler. The high fluctuation in the $\gamma$ values visible from (b) and (c), is indicative of how $\gamma$s are a poor indicator in the presence of datagaps.}
\end{figure}

In this section, we consider the variation of the $\Delta\alpha$ = $\alpha_2$-$\alpha_1$, $\gamma_2$ and $\gamma_1$ with varying gap size, $m_s$ and position, $m_p$. As a sample case, we show the calculated $f(\alpha)$ spectrum for the evenly sampled time series and two cases with datagaps in Fig. \ref{fig:falvarross}, for the R{\"o}ssler system. The full variation of $\Delta\alpha$ with $m_s$ and $m_p$ for the R{\"o}ssler system is shown in Fig. \ref{fig:alpvar}, where we plot the relative variation, $\delta \Delta\alpha^{rel}$ =   $\frac{|\Delta\alpha-\Delta\alpha_E|}{\Delta\alpha_E}$. For a fixed value of $m_p$, as we increase $m_s$, the value of $\Delta\alpha$ deviates heavily from the evenly sampled value and reaches a peak at around 1$\tau$. On increasing $m_s$ further, it falls again and saturates at a value close to the evenly sampled value. We find that in the critical region identified, the f($\alpha$) spectrum widens considerably for the R{\"o}ssler system. Since the shape of the curve changes considerably when gaps are present, we find that $\gamma$ values fluctuate heavily. The  variation of $\delta\gamma^{rel}$ = $\frac{|\gamma-\gamma_E|}{\gamma_E}$ values with $m_s$ and $m_p$ is shown in Fig. \ref{fig:gam1ross} and \ref{fig:gam2ross}. The heavy fluctuations in the figure makes it clear that $\gamma$ values derived from a time series contaminated with gaps cannot be trusted, in general.

The analysis was conducted for the Lorenz system as well, where the variation is much smaller in comparison. Hence we conclude that while the width of the $f(\alpha)$ spectrum may remain broadly unchanged, except at small $m_p$ and for $m_s\approx 1\tau$, the overall shape of the curve changes, as signified by the change in the $\gamma$ values. We point out that widening of the $f(\alpha)$ curve can also happen due to noise contamination\cite{har09mul}.

\subsection*{Surrogate Analysis}
\begin{figure}[hp]
\centering
\begin{subfigure}{\textwidth}
\includegraphics[width=.45\textwidth]{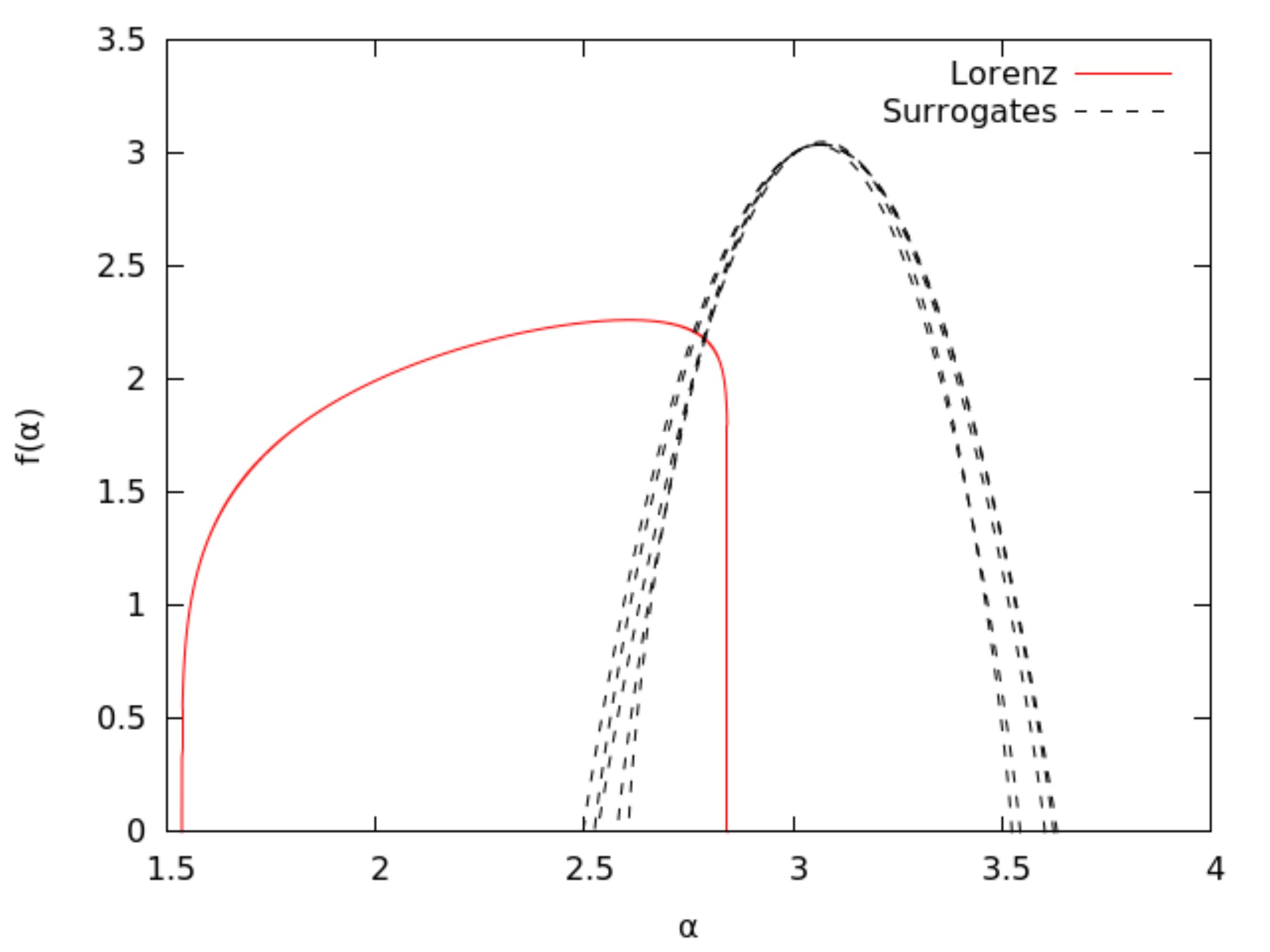}
\caption{\label{fig:fal_surrLor}}
\end{subfigure}
\begin{subfigure}{\textwidth}
\includegraphics[width=.45\textwidth]{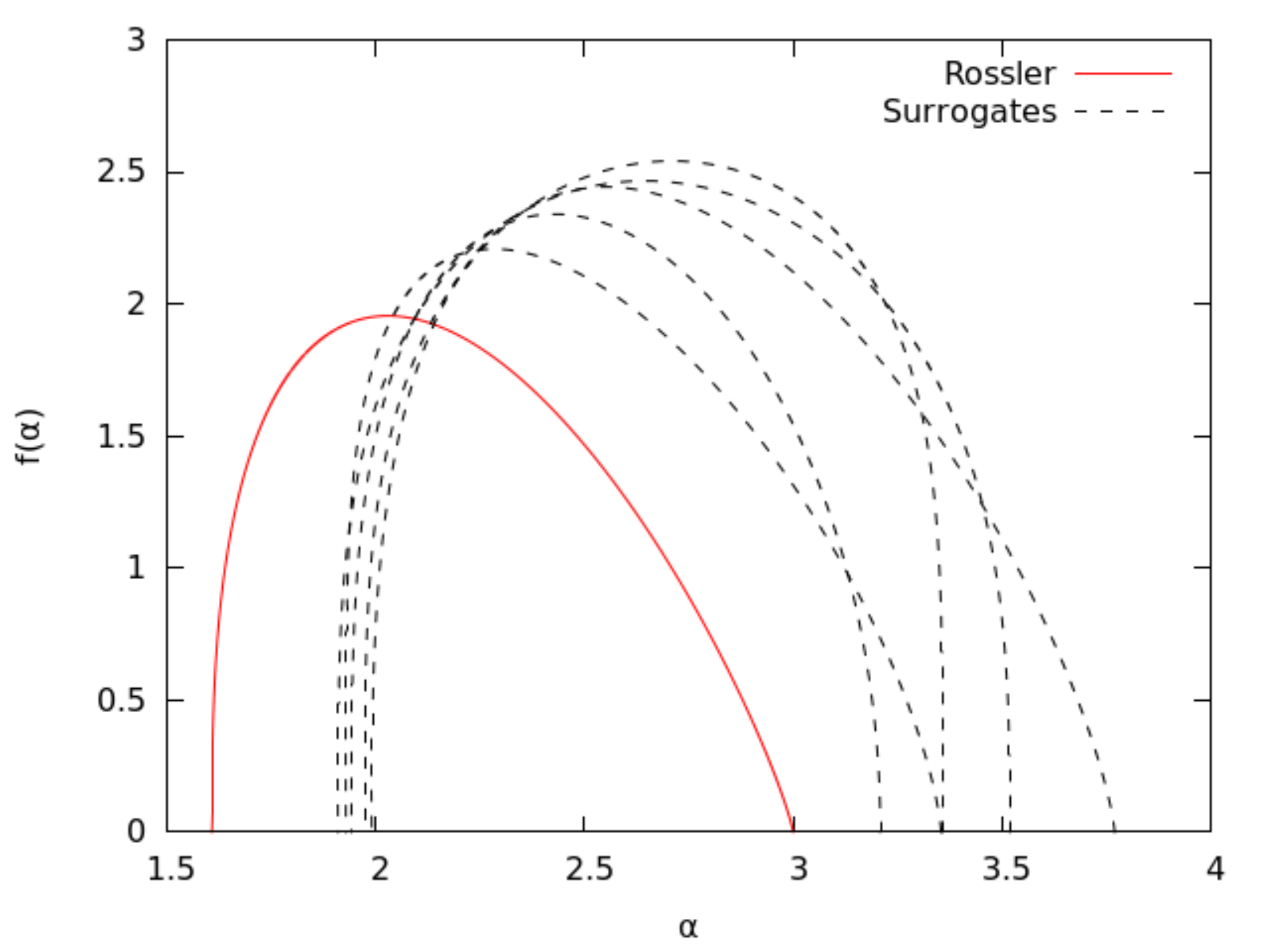}
\caption{\label{fig:fal_surrRN}}
\end{subfigure}
\begin{subfigure}{\textwidth}
\includegraphics[width=.45\textwidth]{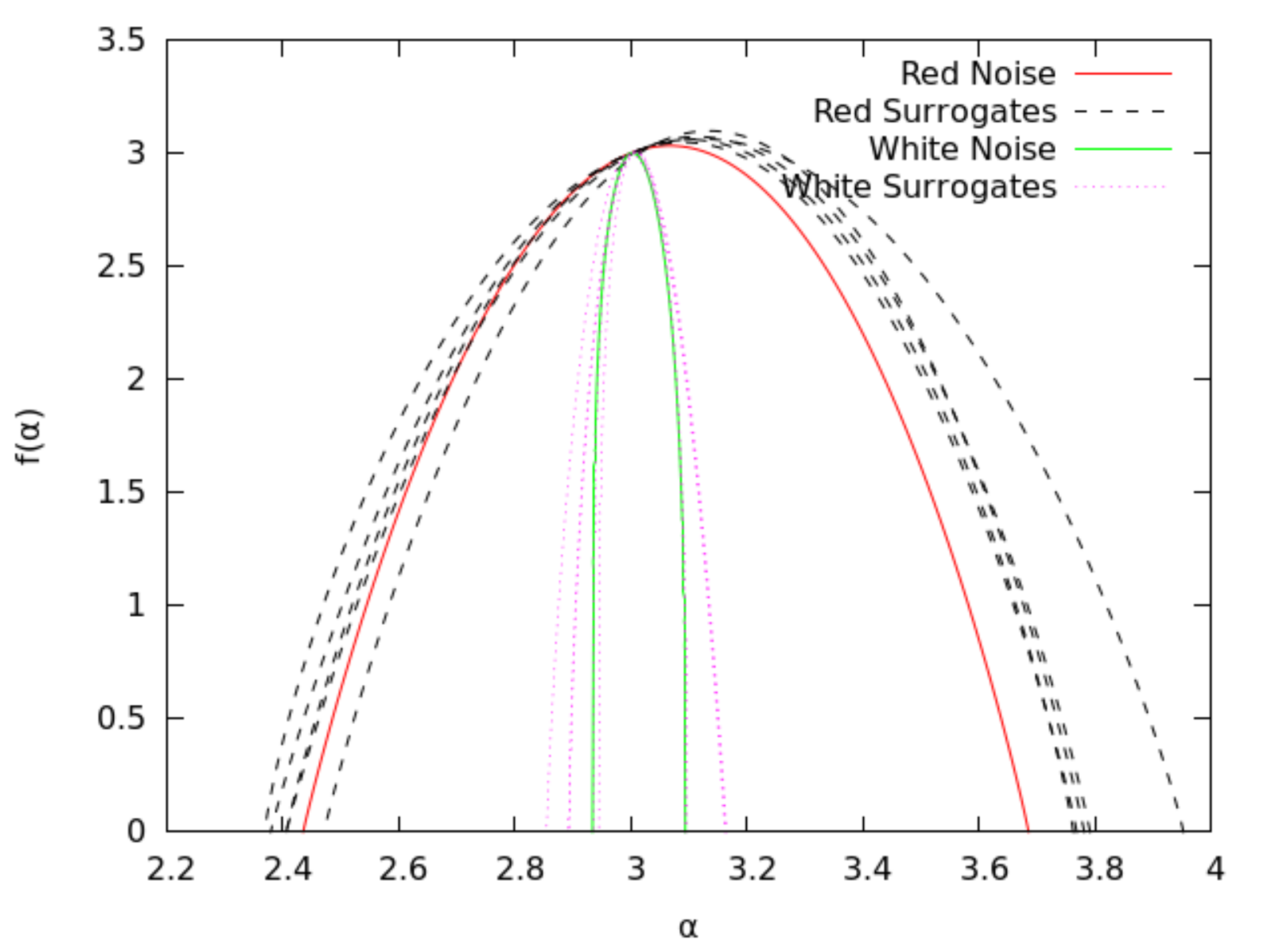}
\caption{\label{fig:fal_surrRN}}
\end{subfigure}

\caption{\label{fig:ES_surr} $f(\alpha)$ vs $\alpha$ curves for (a)Lorenz system (b) R{\"o}ssler system and (c) red and white noise and their surrogates. (a) and (b) show distinct difference between the original data and surrogates, while they are identical for noise.} 
\end{figure}

A wide $f(\alpha)$ spectrum is indicative of multifractality in the system. However this is not conclusive evidence of deterministic nonlinearity. Linear stochastic processes were shown to have saturating $D_2$ and wide multifractal curves in literature\cite{osborne1989finite, harikrishnan2009computing}. A surrogate analysis is required to confirm the existence of multifractality due to deterministic nonlinearity in the system. For this we initially produce 5 Iterative Amplitude Adjusted Fourier Transform (IAAFT) surrogates of the evenly sampled data implemented using the TISEAN package\cite{schreiber1996improved,hegger1999practical}. The $f(\alpha)$ curves obtained from the Lorenz system, R{\"o}ssler system, red noise, white noise and their surrogates are shown in Fig. \ref{fig:ES_surr}. The $\alpha_1$ and $\alpha_2$ values in all the cases studied are tabulated in Table \ref{table:evensamp_surr}. It is clear that for the deterministic systems, the data and surrogates have clearly differing $f(\alpha)$ curves, while for noise, they are not distinguishable.

\begin{table}[hb] 
\caption{$\alpha_1$, $\alpha_2$ values for Lorenz, R{\"o}ssler, Red noise, White noise and their surrogates. $\alpha_1^d$ and $\alpha_2^d$ represent the values for data while $\alpha_1^s$ and $\alpha_2^s$ represent the corresponding values for surrogates.} 
\centering 
\begin{tabular}{c c c c c} 
\hline\hline 
Data  & $\alpha_1^d$ & $\alpha_2^d$ & $\alpha_1^s$ & $\alpha_2^s$\\ [0.5ex] 
\hline 
Lorenz & 1.54 & 2.84 & 2.55$\pm$0.04 & 3.58$\pm$0.04 \\ 
R{\"o}ssler & 1.60 & 2.99 & 1.95$\pm$0.03 & 3.44$\pm$0.19 \\ 
Red Noise & 2.43 & 3.68 & 2.41$\pm$0.04  & 3.81$\pm$0.07\\  
White Noise & 2.93 & 3.09 & 2.90$\pm$0.03  & 3.13$\pm$0.03\\
[1ex] 
\hline 
\end{tabular} 
\label{table:evensamp_surr} 
\end{table}

\begin{figure}[ht]
\centering
\begin{subfigure}{\textwidth}
\includegraphics[width=.45\textwidth]{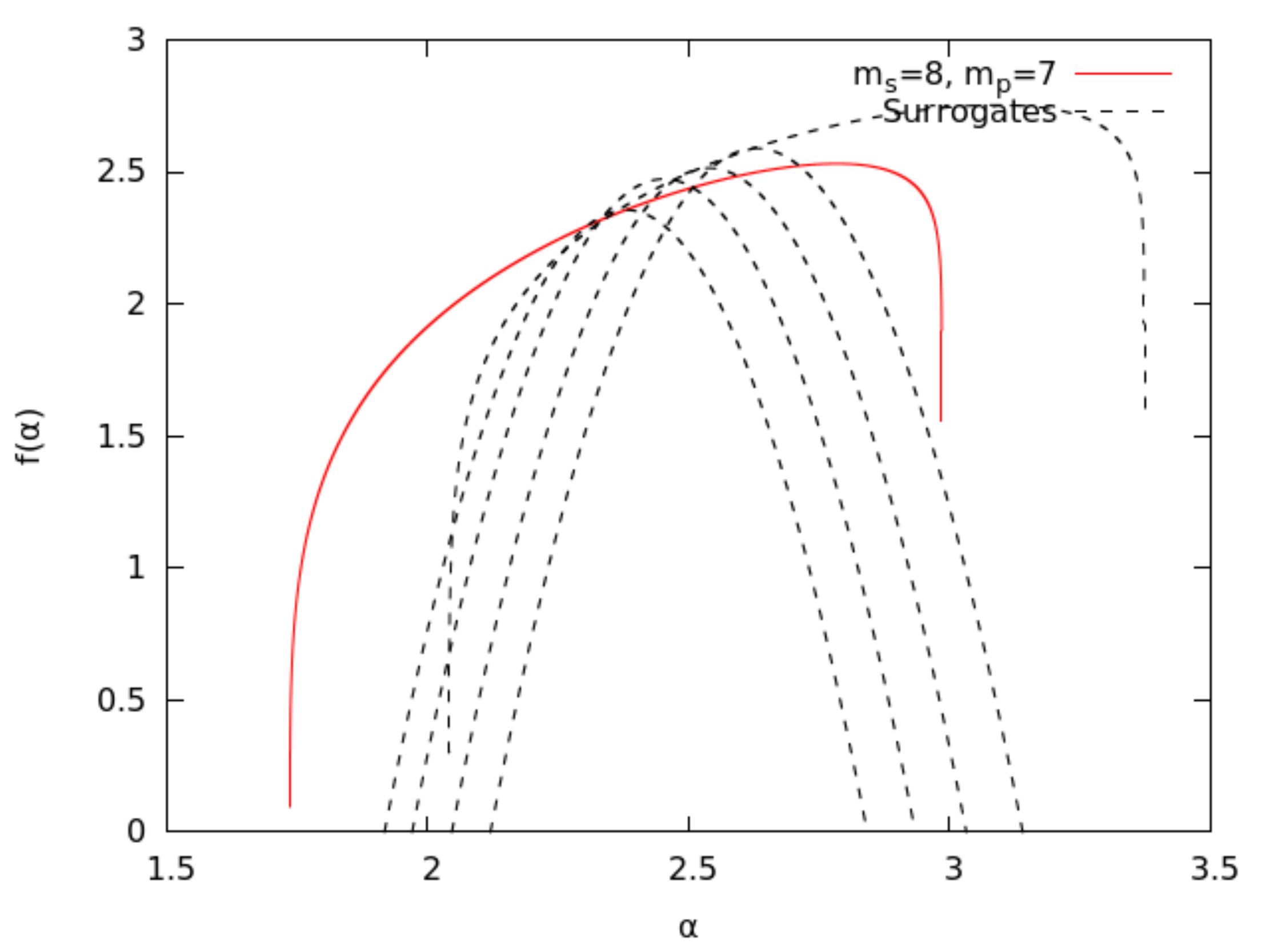}
\caption{\label{fig:fal_surrms8}}
\end{subfigure}
\begin{subfigure}{\textwidth}
\includegraphics[width=.45\textwidth]{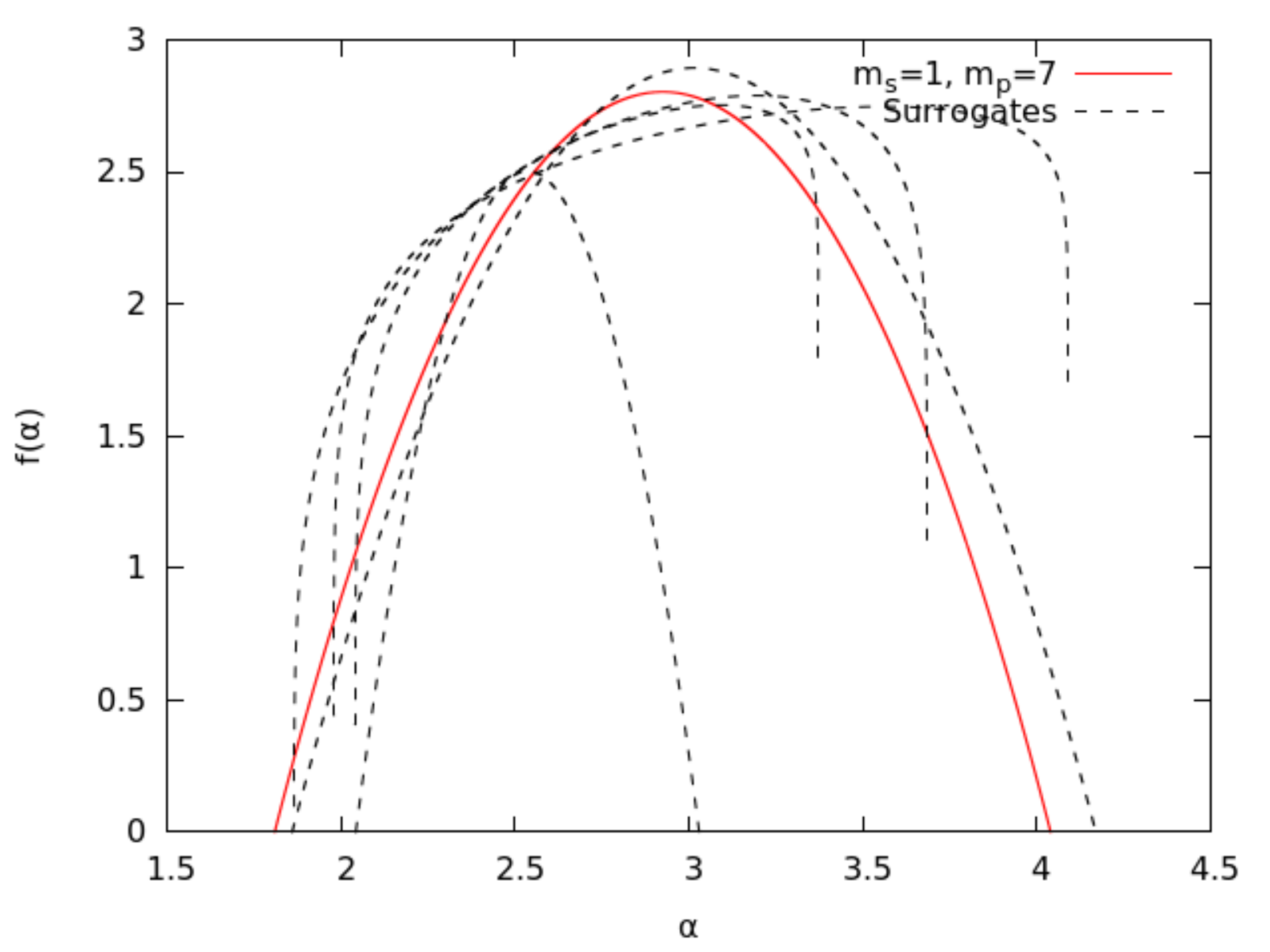}
\caption{\label{fig:fal_surrms1}}
\end{subfigure}
\caption{\label{fig:rosgap_surr} Function fits for $f(\alpha)$ for R{\"o}ssler system and surrogates from (a)$m_s$ = 8$\tau m_p$ = 7 $\tau$ and (b) $m_s$ = 1 $\tau m_p$ = 7 $\tau$. For (a) the data and surrogates can be differentiated, but they merge with each other in (b). (Some curves do not touch the x-axis as the function (Eqn. \ref{eqn:fal}) varies too fast for small $\gamma$ values.)} 
\end{figure}

We repeat the analysis using data with gaps. The same profile of gaps is used to remove data from the surrogates. We find that in the presence of gaps, the original data and surrogates become increasingly difficult to differentiate. In Fig. \ref{fig:rosgap_surr} we show function fits for two cases of the R{\"o}ssler system from a region where the gaps are tolerable and where gaps cause high variation. We see that data merges with surrogates in the region where gaps are not tolerable. This is also evident from Table \ref{table:gaps_surr_ross} which lists the values for $\alpha_1$ and $\alpha_2$ for data and the corresponding values averaged over five surrogate time series. Hence we conclude that in the regions of gap distribution we identify to be non tolerable, multifractality alone cannot confirm deterministic nonlinearity.

\begin{table}[hb] 
\caption{$\alpha_1$, $\alpha_2$ values for R{\"o}ssler and its surrogates for two different values of $m_s$ and $m_p$. We notice that in both cases, $\alpha_1^d$ and $\alpha_1^s$ remains different from each other, whereas $\alpha_2^d$ merges with $\alpha_2^s$.} 
\centering 
\begin{tabular}{c c c c c c} 
\hline\hline 
$m_s$ & $m_p$  & $\alpha_1^d$ & $\alpha_2^d$ & $\alpha_1^s$ & $\alpha_2^s$\\ [0.5ex] 
\hline 
8$\tau$ & 7$\tau$ & 1.73 & 2.98 & 2.02$\pm$0.03 & 3.06$\pm$0.18 \\ 
1$\tau$ & 7$\tau$ & 1.81 & 4.04 & 1.95$\pm$0.07 & 3.82$\pm$0.43 \\ 
[1ex] 
\hline 
\end{tabular} 
\label{table:gaps_surr_ross} 
\end{table}

\section{Analysis of datasets with gaps}
The analysis conducted in the sections above are with the aim of studying variations in $f(\alpha)$ in the presence of datagaps and identifying what the tolerable regions for gap size and frequency are, for the validity of the conclusions drawn from the spectrum to hold. We now apply the analysis to real world datasets. We use datasets with missing data, from the Station for Measuring Forest Ecosystem-Atmosphere Relation (SMEAR), Finland. We look into the time series of air temperature and dew point data measured by the SMEAR II station in the Hyyti{\"a}l{\"a} forest, and datasets of variation of CO$_2$ exchange, photosynthetically active radiation and soil moisture from the SMEAR I station situated in the V{\"a}rri{\"o} forest. 

In all the datasets, we calculate the correlation time, $\tau$ and estimate $m_s$ and $m_p$ in terms of $\tau$. In order to find the dimension used to embed the attractor in, we calculate saturated $D_2$ for the dataset and pick the smallest integer greater than $D_2$ as the embedding dimension. The calculation of $D_2$ filters out all non saturating datasets (i.e. noise processes). Hence the aim of calculating the $f(\alpha)$ spectrum is not just to detect nonlinearity, but to characterize the multifractal properties of the attractor. 

\subsection*{Datasets from SMEAR I}
The SMEAR I station in V{\"a}rri{\"o}, Finland measures a number of meteorological and ecological parameters. The station is situated about 200 km north of the Arctic circle. The datasets produced by the instruments are however not evenly sampled and contain datagaps\cite{zliobaite2014regression}. Some of the primary reasons for gaps in data in this context include thunderstorms and breaks in electricity \cite{hari2012physical}.

In this work we consider the time series of CO$_2$ exchange in pine trees, which acts as a proxy for photosynthesis rate\cite{schneider1941soilmoisturephotosyn} and two factors that affect the photosynthesis rate, namely variation of photosynthetically active radiation on top of the measurement chamber and variation of soil moisture. The rate of photosynthesis has already been suspected of chaotic variation by multiple authors\cite{Luttge1992ChaosCAMphotosyn,Krempaský1993ChaosPhotosynth}. However, to the best of our knowledge no study has been conducted to determine the multifractal characteristics of its embedded phase space structure. Photosynthesis is well known to be dependent on ambient temperature, concentration of atmospheric CO$_2$, amount of photosynthetically active radiation(PAR), soil moisture (SM) etc\cite{Farquhar1980BiochemicalModelPhotosynthesis,shimshi1963soilmoisture}. In this context the study of the nonlinear properties of these variables become important as well. We also study the multifractal properties of the embedded attractors of the time series of variation of photosynthetically active radiation and soil moisture.

All three time series are averaged over half an hour for a period of 12 years starting from 2005 to 2017 and the $m_s$, $m_p$, $\tau$ and $D_2$ values are shown in Table \ref{table:smear1}.
\begin{table}[hb] 
\caption{$\tau$, $m_s$, $m_p$ and $D_2$ for time series of CO$_2$ exchange, Photosynthetically Active Radiation (PAR) and Soil Moisture (SM). } 
\centering 
\begin{tabular}{c c c c c} 
\hline\hline 
Data &$\tau(hours)$ & M$_{s}(hours)$ & M$_{p}(hours)$ & $D_2$\\ [0.5ex] 
\hline 
CO$_2$ Ex. & 5 & 119.6 & 157.3 & 3.18 \\ 
PAR & 7 & 119.6 & 157.3  & 4.08\\ 
SM & 970 & 499.5 &971.1 & 1.15\\
[1ex] 
\hline 
\end{tabular} 
\label{table:smear1} 
\end{table}

\begin{figure}[ht]
\centering
\includegraphics[width=.45\textwidth]{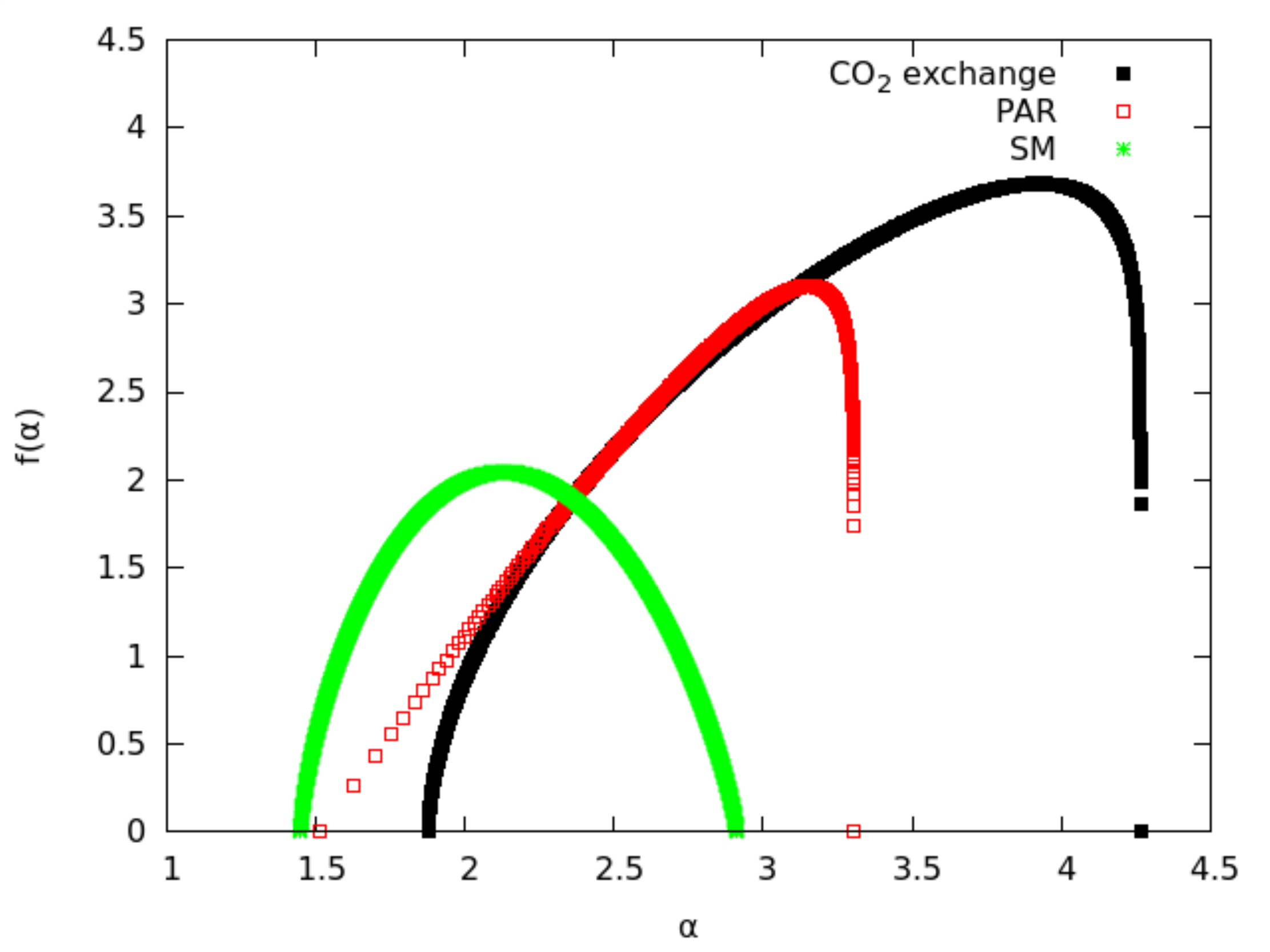}
\caption{\label{fig:smear1falp} $f(\alpha)$ vs $\alpha$ for time series of CO$_2$ exchange in a pine shoot, Photosynthetically Active Radiation (PAR) and  Soil Moisture (SM) from the SMEAR I  station. All three show multifractality. }
\end{figure}

We find that barring the time series for soil moisture, most of the $m_s$ and $m_p$ are outside the identified critical region. Hence in the $f(\alpha)$ computed, $\alpha_1$ and $\alpha_2$ are likely to be largely unaffected. In the case of soil moisture it is possible that the width of the curve in the absence of datagaps, may be much smaller than the one calculated. The plots of the $f(\alpha)$ spectra are shown in Fig. \ref{fig:smear1falp}.
We then proceed to do surrogate analysis to check for signs of deterministic nonlinearity. Since the time series has missing data we cannot directly apply the method of IAAFT mentioned above. Instead we take evenly sampled segments of the data to construct surrogates. We first take segments of data that are continuously sampled. Surrogates are constructed for each segment, and  joined together to act as a surrogate for the whole data. We caution that this may not always be possible and is dependent upon the availability of gap-less chunks in the the dataset. As before we find the $\alpha_1$ and $\alpha_2$ values for data and surrogates, presented in Table \ref{table:smear1_surr}. We find that for CO$_2$ exchange and photosynthetically active radiation have values of  $\alpha_1$ and $\alpha_2$ distinctly different from their surrogates, suggesting that multifractality due to underlying deterministic nonlinearity can be concluded. This adds weight to the claims in literature that photosynthesis may be varying chaotically\cite{Krempaský1993ChaosPhotosynth,Luttge1992ChaosCAMphotosyn}. No such conclusion can be drawn for the variation of soil moisture, as the data and surrogates lie very close to each other. This could be as a result of the value of $m_s$ and $m_p$ lying very close to the identified critical range or due to a lack of deterministic nonlinearity or noise contamination.
\begin{table}[hb] 
\caption{$\alpha_1$, $\alpha_2$ values for CO$_2$ exchange, photosynthetically active radiation, soil moisture and their surrogates. $\alpha_1^d$ and $\alpha_2^d$ are distinctly different from $\alpha_1^s$ and $\alpha_2^s$ for CO$_2$  exchange and PAR, while they are within errors of each other for SM. This could be either because soil moisture has no underlying deterministic nonlinearity, or because of the value of $m_s$ and $m_p$ lying in the identified critical region.} 
\centering 
\begin{tabular}{c c c c c} 
\hline\hline 
Data  & $\alpha_1^d$ & $\alpha_2^d$ & $\alpha_1^s$ & $\alpha_2^s$\\ [0.5ex] 
\hline 
CO$_2$ Ex. & 1.87 & 2.26 & 2.55$\pm$0.01 & 4.75$\pm$0.01 \\ 
PAR & 2.10 & 3.08 & 3.39$\pm$0.07 & 3.88$\pm$0.09 \\ 
SM & 1.45 & 2.91 & 1.35$\pm$0.02  & 2.88$\pm$0.03\\  
[1ex] 
\hline 
\end{tabular} 
\label{table:smear1_surr} 
\end{table}

\subsection*{Datasets from SMEAR II}
\begin{figure}[H]
\centering
\includegraphics[width=.45\textwidth]{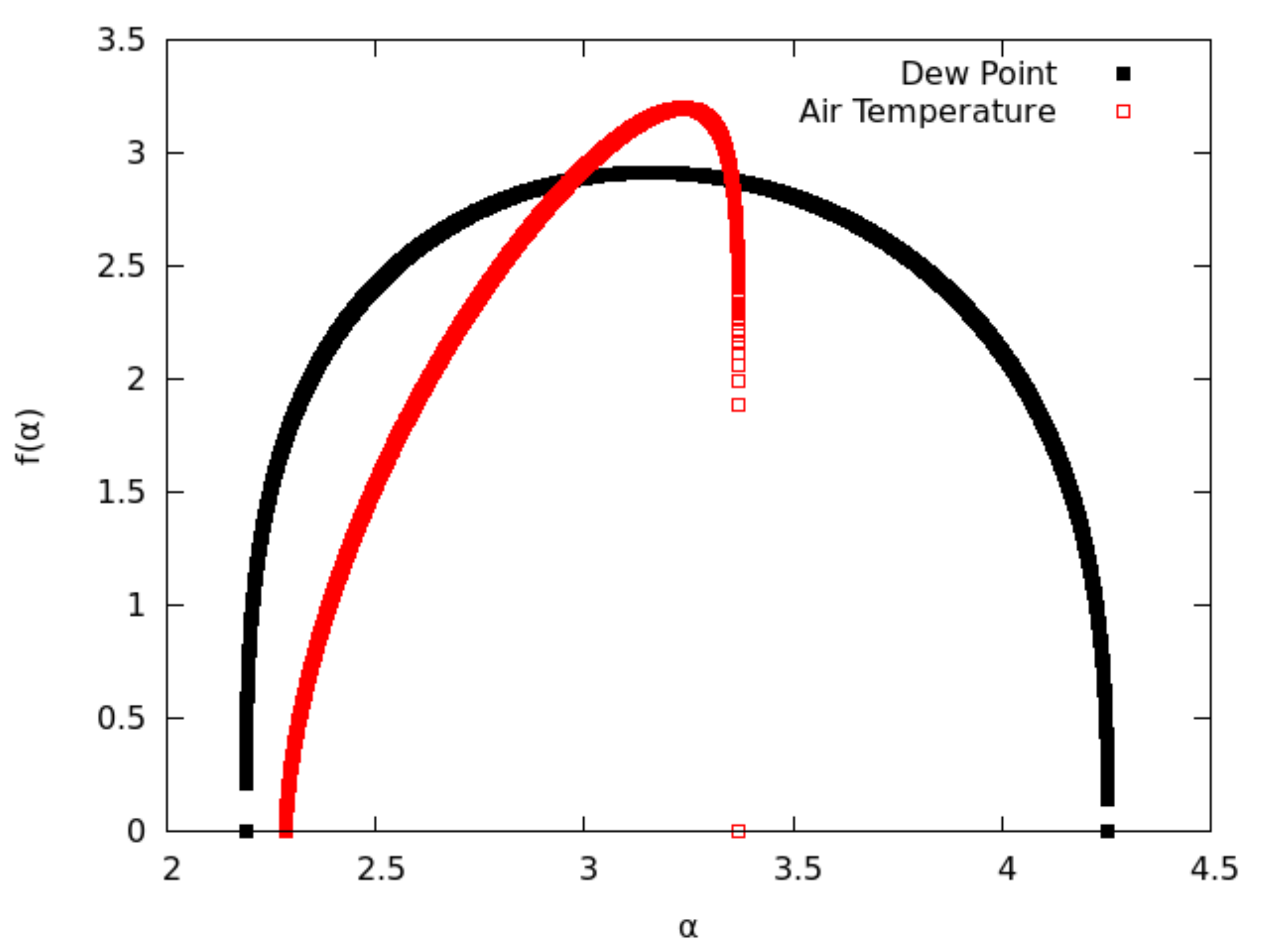}
\caption{\label{fig:smear2falp} $f(\alpha)$ vs $\alpha$ for time series of dew point and air temperature from the SMEAR II  station. Both datasets show multifractality. }
\end{figure}

The SMEAR II station is  established in the Hyyti{\"a}l{\"a} forest in Finland and measures atmospheric aerosols, eco-physiology, soil and water measurements, solar and terrestrial radiation and meteorological measurements. In this work we primarily consider two meteorological datasets, the dew-point and air temperature. These were chosen partially due to the high instance of datagaps in them\cite{zliobaite2014regression}. Climate time series have been often subject to studies of multifractality, in the past\cite{de1999multifractal,kavasseri2005multifractal}. We consider the data sampled every half hour over the period from 2008-2017.
As before we first determine the $m_s$, $m_p$, $\tau$ and $D_2$, shown in Table \ref{table:smear2}.
\begin{table}[ht] 
\caption{$\tau$, $m_s$, $m_p$ and $D_2$ for time series of dew point and air temperature variation. } 
\begin{tabular}{c c c c c} 
\hline\hline 
Data &$\tau(hours)$ & M$_{s}(hours)$ & M$_{p}(hours)$ & $D_2$\\  
\hline 
Dew Pt. & 1000 & 219.2 & 252.7 & 2.63 \\ 
Air T. & 667 & 483.2 & 535.1  & 3.28\\ 
\hline 
\end{tabular} 
\label{table:smear2} 
\end{table}
We notice that $m_s$ for the air temperature is in the critical region identified and hence the $f(\alpha)$ curve may be significantly affected. The falpha curves for both are plotted in Fig. \ref{fig:smear2falp}.

With surrogates constructed as before, we perform a surrogate analysis for datasets from SMEAR II. The results are presented in Table \ref{table:smear2_surr}. For the dew point variation data, both $\alpha_1$ and $\alpha_2$ lie close to the value for the surrogate datasets. Since the $m_s$ and $m_p$ values are away from the identified critical region, it appears that dew point variation has no underlying deterministic nonlinearity and is probably a linear stochastic process. The air temperature time series has $\alpha_1$ and $\alpha_2$ values close to the surrogates, but not within error bars. This could be as a result of the value of $m_s$ and $m_p$ lying in the identified critical region.

\begin{table}[ht] 
\caption{$\alpha_1$, $\alpha_2$ values for dew point and air temperature variation time series. $\alpha_1^d$ and $\alpha_2^d$ are close to $\alpha_1^s$ and $\alpha_2^s$ for both time series. In case of dew point, we suspect this could be as a result of no underlying deterministic nonlinearity. However  for air temperature the value of $m_s$ and $m_p$ lying in the identified critical region, may also be responsible.} 
\centering 
\begin{tabular}{c c c c c} 
\hline\hline 
Data  & $\alpha_1^d$ & $\alpha_2^d$ & $\alpha_1^s$ & $\alpha_2^s$\\ [0.5ex] 
\hline 
Dew Pt. & 2.19 & 4.25 & 2.31$\pm$0.06 & 4.25$\pm$0.05 \\ 
Air T. & 2.29 & 3.79 & 1.95$\pm$0.01 & 4.16$\pm$0.04 \\ 
[1ex] 
\hline 
\end{tabular} 
\label{table:smear2_surr} 
\end{table}

\section{Conclusion}
We investigate the effect of gaps in data on the  $f(\alpha)$ spectrum of the reconstructed attractor. We use data from two standard systems, the R{\"o}ssler and Lorenz systems, to check the effect of datagaps on the computed multifractal spectrum. From the analysis on standard chaotic data we could arrive at tolerable limits on gap parameters, within which conclusions from nonlinear time series analysis can be meaningful. Since these limits are linked to the embedding technique itself, we take that the same can be valid for real world datasets analyzed using the same procedure of embedding. In this work, we illustrate this by analyzing 5 ecological and meteorological datasets with gaps of varying ranges. We support our conclusions by carrying out detailed surrogate analysis on all of them.

We specifically study the variations in the four quantifiers of $f(\alpha)$, viz. $\alpha_1$, $\alpha_2$, $\gamma_1$ and $\gamma_2$, in all the datasets. To the best of our knowledge, our work for the first time establishes multifractality due to deterministic nonlinearity in photosynthesis variation and factors affecting it.

The absence of large evenly sampled datasets is a major impediment for calculating measures applying nonlinear time series analysis techniques and especially deriving quantifiers from $f(\alpha)$. The dangers of interpolation makes it an undesirable solution in these cases. Hence it becomes important to consider how much deviation from the evenly sampled case, actually happens if we ignore gaps. We find that certain broad conclusions about the data can still be retained even in the presence of missing data. We suggest the following procedure while dealing with time series data contaminated with gaps. The means of the gap distributions, $m_s$ and $m_p$ should initially be quantified. If they are not within tolerable limits, binning can be suggested as a method to change the values of $m_s$ and $m_p$ and bring them into the tolerable range identified\cite{geo15}. Once $m_s$ and $m_p$ are within tolerable ranges, reliable conclusions can be drawn about $\alpha_1$ and $\alpha_2$. For any conclusions about underlying deterministic nonlinearity, one needs to conduct an analysis for comparison with phase randomized surrogates. If the surrogates give very different values for quantifiers, from the values for data, one can conclude the existence of an underlying nonlinearity. If surrogates and data merge within tolerable range, the underlying process can be assumed to be purely stochastic or deterministic with high noise contamination. In cases where, even with considerable binning the $m_s$ and $m_p$ values cannot be brought to tolerable ranges, no conclusions can be drawn based on multifractal analysis with confidence.


\section*{Acknowledgements}
 We acknowledge with thanks the data from the SMEAR database that are used in this research. SVG thanks Ron Sunny, IISER, Pune for help with the SMEAR datasets.




\begin{thebibliography}{10}

\bibitem{gra86}
Peter Grassberger.
\newblock Do climatic attractors exist?
\newblock {\em Nature}, 323(6089):609--612, 1986.

\bibitem{geo15}
Sandip~V George, G~Ambika, and R~Misra.
\newblock {Effect of data gaps on correlation dimension computed from light
  curves of variable stars}.
\newblock {\em Astrophysics and Space Science}, 360(1):1--11, 2015.

\bibitem{buc95}
J~Robert Buchler, Thierry Serre, Zolt{\'a}n Koll{\'a}th, and Janet Mattei.
\newblock {A chaotic pulsating star: The case of R Scuti}.
\newblock {\em Physical Review Letters}, 74(6):842, 1995.

\bibitem{zliobaite2014regression}
I~{\v{Z}}liobait{\.e}, Jaakko Hollm{\'e}n, and Heikki Junninen.
\newblock Regression models tolerant to massively missing data: a case study in
  solar-radiation nowcasting.
\newblock {\em Atmospheric Measurement Techniques}, 7(12):4387--4399, 2014.

\bibitem{eckmann1992LimitationOndDimensions}
J-P Eckmann and David Ruelle.
\newblock Fundamental limitations for estimating dimensions and lyapunov
  exponents in dynamical systems.
\newblock {\em Physica D: Nonlinear Phenomena}, 56(2-3):185--187, 1992.

\bibitem{jeffries1998bicoherenceElectrical}
WQ~Jeffries, JA~Chambers, and DG~Infield.
\newblock Experience with bicoherence of electrical power for condition
  monitoring of wind turbine blades.
\newblock {\em IEE Proceedings-vision, image and signal processing},
  145(3):141--148, 1998.

\bibitem{geo17}
Sandip~V George, G~Ambika, and R~Misra.
\newblock Detecting dynamical states from noisy time series using bicoherence.
\newblock {\em Nonlinear Dynamics}, 2017.

\bibitem{hil00}
{Hilborn, Robert C}.
\newblock {\em {Chaos and nonlinear dynamics: an introduction for scientists
  and engineers}}.
\newblock Oxford University Press, 2000.

\bibitem{har09mul}
KP~Harikrishnan, R~Misra, G~Ambika, and RE~Amritkar.
\newblock Computing the multifractal spectrum from time series: an algorithmic
  approach.
\newblock {\em Chaos: An Interdisciplinary Journal of Nonlinear Science},
  19(4):043129, 2009.

\bibitem{hari2013station}
Pertti Hari, Eero Nikinmaa, Toivo Pohja, Erkki Siivola, Jaana B{\"a}ck, Timo
  Vesala, and Markku Kulmala.
\newblock Station for measuring ecosystem-atmosphere relations: Smear.
\newblock In {\em Physical and Physiological Forest Ecology}, pages 471--487.
  Springer, 2013.

\bibitem{Farquhar1980BiochemicalModelPhotosynthesis}
G.~D. Farquhar, S.~von Caemmerer, and J.~A. Berry.
\newblock A biochemical model of photosynthetic co2 assimilation in leaves of
  c3 species.
\newblock {\em Planta}, 149(1):78--90, 1980.

\bibitem{shimshi1963soilmoisture}
Daniel Shimshi.
\newblock Effect of soil moisture and phenylmercuric acetate upon stomatal
  aperture, transpiration, and photosynthesis.
\newblock {\em Plant physiology}, 38(6):713, 1963.

\bibitem{Mogens1985GlobalUniversalityFalp}
Mogens~H. Jensen, Leo~P. Kadanoff, Albert Libchaber, Itamar Procaccia, and Joel
  Stavans.
\newblock Global universality at the onset of chaos: Results of a forced
  rayleigh-b\'enard experiment.
\newblock {\em Phys. Rev. Lett.}, 55:2798--2801, Dec 1985.

\bibitem{de1999multifractal}
MIP De~Lima and J~Grasman.
\newblock Multifractal analysis of 15-min and daily rainfall from a semi-arid
  region in portugal.
\newblock {\em Journal of hydrology}, 220(1):1--11, 1999.

\bibitem{ivanov1999multifractality}
Plamen~Ch Ivanov, Luis A~Nunes Amaral, Ary~L Goldberger, Shlomo Havlin,
  Michael~G Rosenblum, Zbigniew~R Struzik, and H~Eugene Stanley.
\newblock Multifractality in human heartbeat dynamics.
\newblock {\em Nature}, 399(6735):461--465, 1999.

\bibitem{kavasseri2005multifractal}
Rajesh~G Kavasseri and Radhakrishnan Nagarajan.
\newblock A multifractal description of wind speed records.
\newblock {\em Chaos, Solitons \& Fractals}, 24(1):165--173, 2005.

\bibitem{harikrishnan2011nonlinear}
KP~Harikrishnan, Ranjeev Misra, and G~Ambika.
\newblock Nonlinear time series analysis of the light curves from the black
  hole system grs1915+ 105.
\newblock {\em Research in Astronomy and Astrophysics}, 11(1):71, 2011.

\bibitem{osborne1989finite}
A~R Osborne and A~Provenzale.
\newblock Finite correlation dimension for stochastic systems with power-law
  spectra.
\newblock {\em Physica D: Nonlinear Phenomena}, 35(3):357--381, 1989.

\bibitem{harikrishnan2009computing}
KP~Harikrishnan, R~Misra, G~Ambika, and RE~Amritkar.
\newblock Computing the multifractal spectrum from time series: an algorithmic
  approach.
\newblock {\em Chaos: An Interdisciplinary Journal of Nonlinear Science},
  19(4):043129, 2009.

\bibitem{schreiber1996improved}
Thomas Schreiber and Andreas Schmitz.
\newblock Improved surrogate data for nonlinearity tests.
\newblock {\em Physical Review Letters}, 77(4):635, 1996.

\bibitem{hegger1999practical}
Rainer Hegger, Holger Kantz, and Thomas Schreiber.
\newblock Practical implementation of nonlinear time series methods: The tisean
  package.
\newblock {\em Chaos: An Interdisciplinary Journal of Nonlinear Science},
  9(2):413--435, 1999.

\bibitem{hari2012physical}
Pertti Hari, Kari Heli{\"o}vaara, and Liisa Kulmala.
\newblock {\em Physical and physiological forest ecology}.
\newblock Springer Science \& Business Media, 2012.

\bibitem{schneider1941soilmoisturephotosyn}
G~William Schneider and NF~Childers.
\newblock Influence of soil moisture on photosynthesis, respiration, and
  transpiration of apple leaves.
\newblock {\em Plant physiology}, 16(3):565, 1941.

\bibitem{Luttge1992ChaosCAMphotosyn}
Ulrich L{\"u}ttge and Friedrich Beck.
\newblock Endogenous rhythms and chaos in crassulacean acid metabolism.
\newblock {\em Planta}, 188(1):28--38, 1992.

\bibitem{Krempaský1993ChaosPhotosynth}
J.~Krempask{\'y}, M.~Smr{\v{c}}inov{\'a}, and P.~Ballo.
\newblock Periodicity and chaos in a photosynthetic system.
\newblock {\em Photosynthesis Research}, 37(2):159--164, 1993.

\end{thebibliography}
\end{document}